\documentclass{ws-ijmpa}
\usepackage{graphicx}

\begin{document}

\newcommand{\SmY}{{m^2_{Y^{++}}}}
\newcommand{\SmX}{{m^2_{Y^{+}}}}
\newcommand{\mY}{{m_{Y^{++}}}}
\newcommand{\mX}{{m_{Y^{+}}}}
\newcommand{\X}{{Y^{+}}}
\newcommand{\Y}{{Y^{++}}}
\newcommand{\suu}{${SU_c(3)} \times
{SU(3)_L} \times {U(1)_X}$}
\newcommand{\sm}{${SU_c(3)} \times
{SU(2)_L} \times {U(1)_Y}$~}
\newcommand{\zd}{$Z\to \ell^{\pm}_i \ell^{\mp}_j$}
\newcommand{\ld}{$\ell_i\to \ell_j\gamma$}
\newcommand{\ldt}{$\ell_i^-\to  \ell_j^+ \ell_k^- \ell_k^-$}
\newcommand{\tm}{331 model}


\title{Lepton flavor violating decay \zd~ in the \tm}
\author{I. Cort\'es-Maldonado}
\address{Facultad de Ciencias
F\'\i sico Matem\'aticas, Benem\'erita Universidad Aut\'onoma de
Puebla, Apartado Postal 1152, Puebla, Pue., M\'exico}
\author{A. Moyotl}
\address{Instituto
de F\'{\i}sica, Universidad Aut\'onoma de Puebla, Apartado Postal
J-48, 72570, Puebla, M\'exico}
\author{G. Tavares--Velasco}
\address{Facultad de Ciencias
F\'\i sico Matem\'aticas, Benem\'erita Universidad Aut\'onoma de
Puebla, Apartado Postal 1152, Puebla, Pue., M\'exico}

\date{\today}
 \maketitle

\begin{abstract}
We study the lepton flavor violating (LFV) decays \zd~ ($\ell_{i,j}=e,\,\mu,\,\tau$) in the
framework of the minimal 331 model. The main contributions arise at the one-loop level via a doubly
charged bilepton with general LFV couplings. We obtain an estimate for the corresponding
branching ratios by using the bounds on the LFV couplings of the doubly charged bilepton from
the
current experimental limits on the decays \ld~ and \ldt. A bound on the bilepton mass is also
obtained  through the current limit on the
anomalous magnetic moment of the muon. It is found that the bilepton 
contributions to LFV $Z$ decays are not expected to be at the reach of experimental detection.
In particular,
the branching ratio for the $Z\to \mu^\pm \tau^\mp$ decay is below the $10^{-10}$ level for a
bilepton mass of the order of 500 GeV.
\end{abstract}

\ccode{PACS number(s): 13.38Dg, 13.35.-r}

\section{Introduction}
In the standard model (SM), neutrinos are considered massless and thus lepton
flavor violation (LFV) is forbidden at any order of perturbation theory.  Even
if the theory is extended with massive neutrinos, LFV transitions such as \ld~
would be induced up to the one-loop level and would be strongly suppressed due
to a GIM-like mechanism: it was found that BR$(\mu \to e\gamma)\simeq
10^{-25}-10^{-45}$ in the SM extended with non-diagonal lepton flavor couplings
and massive neutrinos with a mass $m_\nu$ of a few eVs.\cite{Petcov:1976ff} Any
signal of LFV would thus be a hint of new physics. However, recent evidences of
neutrino oscillations and thereby a nonzero neutrino mass clearly point to LFV
and have thus triggered the interest on the study of LFV decays such as \ld,
\ldt, and \zd~ ($\ell_{i,j}=e,\,\mu,\,\tau$). Several theoretical extensions of the SM
do predict such LFV transitions with a non-negligible rate. Currently there are
stringent experimental constraints on LFV muon decays:\footnote{All the experimental limits
used in this work correspond to the  90\% C.L. limits unless stated otherwise.} BR$(\mu \to e\gamma)
<
2.4 \times 10^{-12}$,\cite{Adam:2011ch} BR$(\mu\to 3e) < 1.0 \times
10^{-12}$,\cite{Bellgardt:1987du} and BR$(\mu {\rm Ti} \to
e {\rm Ti}) < 3.6 \times 10^{-11}$.\cite{Kaulard:1998rb}  Even more, the bound on the $\mu \to
e\gamma$ rate is expected
to be improved up the level of $10^{-13}$ by the MEG experiment.\cite{Ritt:2006cg} On the
other hand, the current bounds on LFV transitions involving the $\tau$ lepton are less stringent:
BR$(\tau\to e \gamma) < 3.3 \times
10^{-8}$,\cite{:2009tk} BR$(\tau\to \mu \gamma) < 4.4 \times
10^{-8}$,\cite{:2009tk} BR$(\tau\to 3e) < 3.6 \times
10^{-8}$,\cite{Aubert:2007pw} and  BR$(\tau\to e^-e^+ \mu) < 3.7 \times
10^{-8}$.\cite{Miyazaki:2007zw} As a matter of fact, the possibility that LFV transitions
involving the last two lepton generations may be larger than those involving the electron has been
widely conjectured in the literature. As far as LFV $Z$ decays are concerned, the most stringent
experimental bounds were obtained at LEP:\cite{Akers:1995gz,Abreu:1996mj}

\begin{eqnarray}
{{\rm BR}}\left(Z \to e^{\mp} \mu^{\pm}\right)&<& 1.7 \times
10^{-6},\\
{{\rm BR}}\left(Z \to e^{\mp} \tau^{\pm}\right)&<& 9.8 \times
10^{-6},\\
{{\rm BR}}\left(Z \to\mu^{\mp} \tau^{\pm}\right)&<&1.2 \times
10^{-5},
\end{eqnarray}
The study of these LFV $Z$ decays has been the source of great interest as the
future international linear collider with its Giga-$Z$ option would allow a
yearly production of $10^9$ $Z$ bosons,\cite{AguilarSaavedra:2001rg} which
would open up the possibility for detecting some
$Z$ boson rare decays. Several predictions for the \zd~ decay width have been
obtained in the framework of various SM extensions,
\cite{Kuo:1985jt,Bernabeu:1987dz,Mendez:1989wg,Korner:1992an,Frank:1996xt,Frank:2000dw,Illana:2000ic,Ghosal:2001ep,Iltan:2001au,FloresTlalpa:2001sp,Yue:2002pk,Cao:2003zv,Iltan:2004af,Iltan:2005jp,Iltan:2008jn,Cao:2009cp} such as supersymmetric theories,
\cite{Mendez:1989wg,Cao:2009cp,Cao:2003zv} the
two-Higgs doublet model,\cite{Iltan:2001au,Iltan:2004af} the Zee-model,
\cite{Ghosal:2001ep} the scalar triplet model,\cite{Bernabeu:1987dz} the
left-right symmetric model,\cite{Frank:1996xt,Frank:2000dw} top-color assisted
technicolor,\cite{Yue:2002pk} the SM with massive neutrinos,
\cite{Kaulard:1998rb,Illana:2000ic} effective theories,
\cite{FloresTlalpa:2001sp}  etc.

The \tm~\cite{Pisano:1991ee,Frampton:1992wt} is an appealing SM extension based
on the $SU(3)_L \times U(1)_X$ gauge group. This model
predicts new physics at the TeV scale and it is also attractive due to its  peculiar mechanism of
anomaly cancellation, which requires that  the fermion family number is
a multiple of the quark color number, thereby suggesting a solution to the flavor
problem.  A remarkable feature of the \tm~is the prediction of new exotic
particles with masses bounded from above at the TeV scale due to
theoretical constraints.  Therefore, such a model could be confirmed or ruled
out in a near future. Among the new particles predicted by the model, there are
singly and doubly charged scalar bosons, exotic quarks of electric charges $-4/3e$ and
$5/3e$, singly and doubly charged gauge bosons, and an extra neutral gauge boson. The new charged
gauge bosons are known as bileptons as they carry two units of lepton number. LFV can
be induced at the tree-level in the scalar and gauge sectors. In particular, we will focus on
the possibility that the new doubly charged gauge bileptons can give rise to LFV at
the tree-level, which in turn can induce LFV $Z$ decays at the one-loop level.
Our aim is to present such a calculation and obtain an estimate for the \zd~
branching ratios. To constrain the LFV bilepton couplings we will use the
current experimental bounds on the  LFV decays \ld~ and \ldt.  The anomalous
magnetic moment of the muon will be used to constrain the bilepton mass.

The rest of our presentation is organized as follows. In Sec. \ref{model} we present an overview of
the minimal \tm~ and consider the possibility of LFV mediated by the doubly charged vector
bilepton. Section. \ref{decay} is devoted to the presentation of our calculation, whereas the
numerical results and the analysis are presented in  Sec. \ref{analysis}. The conclusions and
outlook are presented in Sec. \ref{conclusion}.

\section{The minimal \tm}
\label{model}

The minimal \tm~ is based on the \suu~ gauge group. In this model,
neutrinos are massless and the leptons are accommodated
in antitriplets of $SU(3)_L$:\cite{Pisano:1991ee,Frampton:1992wt}

\begin{equation}
\ell^i_L= \left( \begin{array}{c} e^i_L \\
\nu^i_L \\
e^{c\,i}
\end{array} \right): (1, 3^*,0),
\end{equation}

\noindent where $i=1,2,3$ is the generation index and $e^{c\,i}_L$
is the complex conjugate field of $e^i_L$. Anomaly cancellation
requires that the first two quark generations are represented by
triplets of $SU(3)_L$, while the third one appears as an antitriplet of $SU(3)_L$.
These multiplets are completed by three new exotic quarks ($D$, $S$ and $T$)
with electric charges $Q_{D,S}=-4/3e$ and
$Q_T=5/3e$. For the purpose of this work it is not necessary a
further discussion on the quark sector.

The most economic scalar sector of the minimal \tm~ requires three
scalar triplets and one sextet of ${SU(3)_L}$.\cite{Liu:1993gy} One
scalar triplet is necessary to break ${SU(3)_L}\times {U(1)_X}$ down
to the electroweak gauge group, whereas electroweak symmetry
breaking (EWSB) requires the two remaining scalar  triplets and the
sextet. The latter is necessary to provide realistic masses for the
leptons. This minimal Higgs sector has the following quantum numbers

\begin{equation}
\phi_Y= \left( \begin{array}{c}
\Phi_Y \\
\phi^0
\end{array} \right): \quad (1,3,1); \quad
\phi_1= \left( \begin{array}{c}
\Phi_1 \\
\Delta^-
\end{array} \right): \quad (1,3,0); \quad
\phi_2= \left( \begin{array}{c}
\widetilde{\Phi}_2 \\
\rho^{--}
\end{array} \right): \ \ \ (1,3,-1),
\end{equation}
where $\Phi_i=(\phi^{+}_i,\phi^0_i)$, with
$\widetilde{\Phi}_i=i\,\tau^2\,\Phi^*_i$ for $i=1,\,2,\,3$;
$\Phi_Y=(\Phi_Y^{++},\Phi_Y^+)$ contains the would-be Goldstone bosons
associated with the new doubly charged, $Y^{++}$, and singly charged, $Y^+$,
bileptons; the real and imaginary parts of
$\phi^0$ correspond to one physical Higgs boson and the would-be Goldstone
boson associated with the extra neutral gauge boson, ${Z^\prime}$.
In addition  the scalar sextet  is given by

\begin{equation}
H= \left( \begin{array}{ccc}
T & \frac{\widetilde{\Phi}_3}{\sqrt{2}}\\
\frac{\widetilde{\Phi}^T_3}{\sqrt{2}} & \eta^{--}
\end{array} \right): \ \ \ (1,6,0),
\end{equation}

\noindent where $T$ is a $SU(2)_L$ triplet

\begin{equation}
T= \left( \begin{array}{ccc}
T^{++} & T^+/\sqrt{2}\\
T^+/\sqrt{2} & T^0
\end{array} \right),
\end{equation}
whereas $\Delta^-$, $\rho^{--}$, and $\eta^{--}$ are singlets of
$SU(2)_L$ with hypercharge $-2$, $-4$, and $+4$, respectively.

The covariant derivative in the fundamental representation of
${SU(3)_L}\times {U(1)_X}$ can be written as

\begin{equation}
{D}_\mu=\partial_\mu-i\,g\,\frac{\lambda^a}{2}
\,W^a_\mu-i\,g_XX\,\frac{\lambda^9}{2}\,X_\mu, \qquad (a=1 \ldots
8),
\end{equation}

\noindent with $\lambda^a$ the Gell-man matrices and
$\lambda^9=\sqrt{2/3}\;{\rm diag(1,1,1)}$.  The first stage of
spontaneous symmetry breaking (SSB) is triggered by the  vacuum
expectation value (VEV) of $\phi_Y$, which breaks the
${SU(3)_L}\times {U(1)_X}$ gauge group down to ${SU(2)_L}\times
{U(1)_Y}$. In this stage of SSB,  the exotic quarks and the new gauge
bosons acquire their masses.  The bileptons appear in a ${SU(2)_L}\times
{U(1)_Y}$ doublet with hypercharge $3$ and are mass degenerate. They are defined in terms of the
gauge eigenstates as
follows

\begin{equation}
Y^\mu=\left( \begin{array}{c}
Y^{++}_\mu\\
Y^+_\mu \end{array}\right)=\frac{1}{\sqrt{2}}\left( \begin{array}{c}
W^4_\mu-iW^5_\mu\\
W^6_\mu-iW^7_\mu\end{array}\right).
\end{equation}

The gauge fields $W^8_\mu$ and $X_\mu$ mix to produce the extra
neutral gauge boson, ${Z^\prime}$, along with a massless gauge
boson, $B_\mu$, which is associated with the ${U(1)_Y}$ group.
They are given by
\begin{align}
Z^\prime_\mu&=c_\theta\, W^8_\mu-s_\theta\, X_\mu, \\
B_\mu&=s_\theta W^8_\mu+c_\theta X_\mu,
\end{align}
 where $s_\theta=\sin\theta, c_\theta=\cos\theta$ and
$\tan\theta=g_X/(\sqrt{2}\,g)$. The coupling constant associated
with the hypercharge group is  $g'=g\,s_\theta/\sqrt{3}$. The
remaining fields associated with the unbroken generators of
${SU(3)_L}$ are the gauge bosons of the $SU(2)_L$ group, which are
denoted by $W^i_\mu$ for $i=1,2,3$.

After the first stage of SSB we are left with the SM with its particle
content plus the new  gauge bosons and exotic quarks, together with several scalar multiplets of
$SU(2)_L$: three doublets
$\Phi_i$ ($i=1,\,2,\,3$), one triplet $H$, and various
singlets. Electroweak symmetry breaking (EWSB)
proceeds at the Fermi scale via the VEV of the ${{SU(2)_L}}$ doublets
$<\Phi^0_i>_0=v_i/\sqrt{2}$ ($i=1,2$). By simplicity it can be assumed
that $<H>_0=0$. In this stage, the SM particles acquire their masses and the
bileptons and the ${Z^\prime}$ boson receive additional mass
contributions. The extra mass terms for the bileptons, which arise from
the Higgs kinetic-energy sector, violate the custodial $ SU(2)$ symmetry. Therefore,
the bilepton masses split:

\begin{equation}
\mY=\frac{g^2}{4}\left(u^2+v^2_2\right), \quad
\mX=\frac{g^2}{4}\left(u^2+v^2_1\right). \label{Ymass1a}
\end{equation}
This mass splitting is bounded by the  hierarchy of the SSB:
$|m^2_{Y^+}-m^2_{Y^{++}}|\leq 3m^2_W$. On the other hand, by
matching the gauge coupling constants at the first stage of SSB, it
is found that
\begin{equation}
\frac{g^2_X}{g^2}=\frac{6s^2_W(m_{Z'})}{1-4s^2_W(m_{Z'})},
\end{equation}
which means that $s^2_W(m_{Z'})$ has to be smaller than $1/4$. It
was found that this condition implies that the new $Z'$ boson
cannot be heavier than $3.1$ TeV.\cite{Liu:1993fwa,Ng:1992st} From
this result and the symmetry-breaking hierarchy $u \gg v_1,v_2,v_3$,
it is inferred that the bileptons masses are smaller than
$m_{Z'}/2\simeq 1500$ GeV.

\subsection{LFV in the \tm}
The possibility of LFV in the \tm~ was first analyzed
in Ref. \refcite{Liu:1993gy} and more recently in Ref. \refcite{Blum:2007he} in a more general
context.
In the scalar sector, LFV can be mediated by the neutral and charged scalar
bosons. However, these interactions are expected to be
very suppressed due to the smallness of the Yukawa couplings. We will thus
not consider LFV mediated by scalar bosons in our
calculation.

As far as the gauge sector is concerned,
the neutral $Z'$ gauge boson cannot mediate LFV  at the tree-level: it turns out that
the $Z'$ boson couplings to
the leptons are flavor universal, so this gauge boson cannot mediate LFV at
the tree level as the rotation of flavor states to mass eigenstates
yields a diagonal coupling matrix. Moreover, it is interesting to note that
the $Z'$ gauge boson has a leptophobic nature as its couplings to a
lepton pair are suppressed by the  $\sqrt{1-4 s_W^2}$ factor.
\cite{Perez:2004jc} On the other hand, the interactions between the
bileptons and the leptons can be written in flavor space as

\begin{equation}
{\cal L}=-\frac{g}{\sqrt{2}}\overline{\ell^{'c}_R} \gamma^\mu
\ell'_L Y^{++}_\mu +\frac{g}{\sqrt{2}}\overline{\ell^{'c}_R}
\gamma^\mu \nu'_L Y^{+}_\mu+{\rm H.c.}
\end{equation}
After a rotation to the physical states is performed ($\ell'_{L,R}=U_{L,R} \ell_{L,R}$ and
$\nu'_L=U_L \nu_L$) we are left
with

\begin{equation}
\label{Ylilj}
{\cal L}=-\frac{g}{\sqrt{2}}\overline{\ell^c_R} \gamma^\mu V^Y
\ell_L Y^{++}_\mu +\frac{g}{\sqrt{2}}\overline{\ell^c_R} \gamma^\mu
V^Y \nu_L Y^{+}_\mu+{\rm H.c.}
\end{equation}
where we introduced the unitary flavor mixing matrix $V^Y=U_R^T U_L$. The  singly charged bilepton
effects on the \zd~ decay will vanish due to the zero mass of neutrinos. We will thus only need to
consider LFV mediated by the
doubly charged bilepton.

\section{Analysis of the \zd~decay}
\label{decay}

\begin{figure}
 \centering
\includegraphics[width=4in]{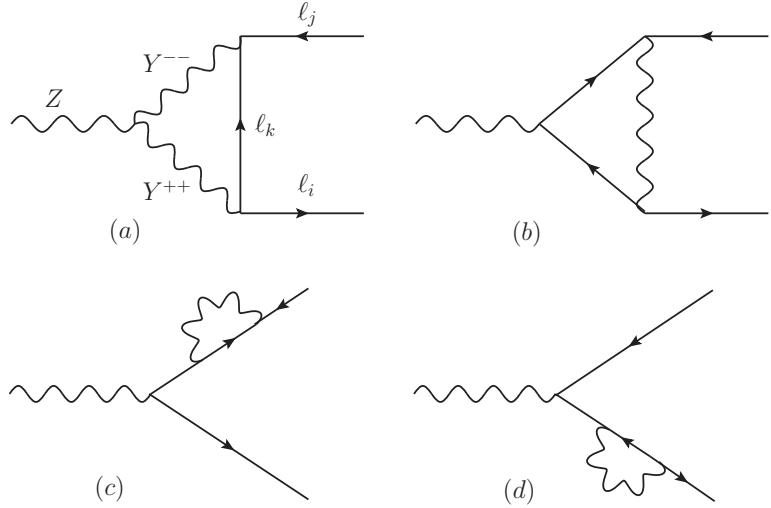}
\caption{\label{fig_fermdiag} Feynman diagrams for the \zd~ decay at the one-loop level in
the \tm. In the Feynman-t'Hooft gauge there is an additional set of
Feynman diagrams obtained by replacing each bilepton gauge boson by
its  associated would-be Goldstone boson, but these diagrams can be neglected since their amplitudes
are proportional to the lepton masses.}
\end{figure}

To calculate the \zd~decay we will neglect the masses of the outgoing leptons but the internal
lepton mass will be retained. The transition amplitude will be calculated using
the Feynman-t'Hooft gauge. In our approximation, the contributions to the
\zd~decay arise from the four Feynman diagrams
shown in Fig. \ref{fig_fermdiag}. Although there is also a set of Feynman diagrams obtained by
replacing each bilepton  by its associated would-be Goldstone
boson, $G_Y$, the respective amplitudes contain terms  proportional to products of the lepton masses
due to the form of the $G_Y \ell_i\ell_j$ coupling, so these contributions can be neglected from the
calculation. Apart from the
Feynman rule for the coupling of the doubly charged bilepton
to a lepton pair, which can be extracted from Eq. (\ref{Ylilj}), we only need the Feynman rule for
the
$ZY^{++}Y^{--}$ vertex. The latter is given in the Feynman-t'Hooft
gauge by:\cite{Montano:2005gs}

\begin{eqnarray}
Z_\alpha(k) Y^{--}_\mu(k_1)
Y^{++}_\nu(k_2)&=&-\frac{ig}{2c_W}g_{ZYY}\Big((k_2-k_1)_\alpha
g_{\mu \nu}+(k-k_2- k_1)_\mu g_{\alpha \nu}\nonumber\\&-&(k-k_1- k_2)_\nu
g_{\alpha \mu}\Big),
\end{eqnarray}
where all the particles are incoming and $g_{ZYY}=1-4s_W^2$, with
$s_W=\sin\theta_W$ and $c_W=\cos\theta_W$.

To obtain the transition amplitude for each Feynman diagram we used the method of Ref.
\refcite{Denner:1992vza}, which is meant for processes involving vertices with complex conjugate
fields, such as our lepton number violating vertices. The Feynman parameters
technique was used to solve the loop integrals. In the massless outgoing lepton limit,
the $Z(p)\to \ell^\pm_i(p_i) \ell^\mp_j(pj)$ decay amplitude
can be written as
\begin{equation}
\label{mzlg}
 i{\cal M}\left(Z\to \ell^\pm_i \ell^\mp_j\right)= F_{L}^{ij}\bar{u}(p_i)
\gamma_\alpha  P_L v(p_j)\epsilon^\alpha(p),
\end{equation}
with $P_{L}=(1- \gamma^5)/2$. The $F^{ij}_{L}$ function depends on the internal lepton mass and the
bilepton mass.  It can be expressed in the form:

\begin{equation}
F_L^{ij}(x_k,x_Y)=\frac{g^3}{2^6\pi^2\,c_W}\sum_{k=1}^3 V^Y_{ik}
V^{Y\,*}_{jk}I(x_k,x_Y), \label{FL}
\end{equation}
with
\begin{equation}
I(x_k,x_Y)=\sum_{n=1}^4 f_n(x_{k},x_Y). \label{Sum}
\end{equation}
We have introduced the notation $x_k=m^2_k/m^2_Z$ and
$x_Y=m^2_Y/m_Z^2$, with $m_k$ the internal lepton mass and $m_Y$ the bilepton mass. The $f_n$
function stands for the contribution of the $n$th Feynman
diagram of Fig. \ref{fig_fermdiag}. After some lengthy algebra, we obtain:

\begin{eqnarray}
f_1(x_k,x_Y)&=&2g_{ZYY}\Big[
\left(   \delta_{Yk}+2 \right)\left(B_{Y}-B_{kY}\right)-\left(\delta_{Yk}^2+ 2x_Y-x_k
\right)C_{Yk}\nonumber\\&-&\frac{1}{2}B_{Y}\Big],\\
\label{f1ft}
f_2(x_k,x_Y)&=&2g_R\left[1+\left(\delta_{Yk}+2\right)\left(B_{k}-B_{kY}\right) - \frac{1}{2}B_{k} +
{\left( 1+ \delta_{Yk} \right) }^2C_{kY}\right]\nonumber\\&+&2g_L x_k C_{kY},\\
\label{f2ft}
(f_3+f_4)(x_k,x_Y)&=& g_L\left(1-B_{kY}-dB_{kY}\right),
\label{f34ft1}
\end{eqnarray}
where  $g_{L,R}=g_V\pm g_A$, with $g_V=-1/2+2s_W^2$ and
$g_A=-1/2$  the vector and axial $Z$-lepton couplings. Note that  the amplitudes of the two bubble
diagrams must be combined
to obtain the right limit for massless outgoing leptons. In addition, the following functions (the
subscript denotes the corresponding dependence) have been introduced:

\begin{eqnarray}
B_{Y}&=&2\left(1 - \tau_Y\,{\rm arccot}\tau_Y - \log(x_Y)\right)+\Delta,\\
\nonumber\\
B_{k}&=&2\left(1 - \lambda_k\,{\rm arccoth}\lambda_k - \log(x_k)\right)+\Delta,\\
\nonumber\\
B_{kY}&=&1 + \frac{1}{\delta_{Yk}}\left(x_k\log(x_k)-x_Y \log(x_Y)\right)+\Delta,\\
\nonumber\\
dB_{kY}&=&\frac{1}{2 {\delta_{Yk}}^3}\left(x_Y^2 -x_k^2  - 2 x_k x_Y
\log\left(\frac{x_k}{x_Y}\right)\right),\\
\nonumber\\
C_{kY}&=&\log\left(\frac{\delta_{Yk}}{\delta_{Yk}+1}\right)\log(\delta_{Yk})+
F\left(\frac{x_Y}{\delta_{Yk}}\right)-F(\lambda_+)-F(\lambda_-),\\
\nonumber\\
C_{Yk}&=&G(x_k)-G(x_Y)-4 \,{\rm arccot}\left({\tau_Y}\right) {\rm
arctan}\left(\frac{\tau_Y}{2\delta_{Yk}-1}\right)\nonumber\\&-&\log\left(\frac{\delta_{Yk}}{
\delta_{Yk}-1}\right) \log\left(x_Y\right)+2\,{\rm Re}\left[H(\tau_+)\right],
\end{eqnarray}
where $\delta_{Yk}=x_Y-x_k$, $\lambda_{\pm}=\frac{1}{2} \left(1 \pm \lambda_k\right)$,
$\lambda_k=\sqrt{1-4 x_k}$,  $\tau_+=\frac{1}{2}\left(1+ i\tau_Y\right)$ and $\tau_Y=\sqrt{4
x_Y-1}$.
$\Delta$ stands for the usual ultraviolet singularity in dimensional regularization. Furthermore
\begin{eqnarray}
F(\lambda)&=&\log (\lambda) \log
\left(\frac{\delta_{Yk}}{\delta_{Yk}+\lambda}\right)-\log(\lambda-1) \log
\left(\frac{\delta_{Yk}+1}{\delta_{Yk}+\lambda}\right)+\text{Li}_2\left(\frac{\lambda}
{\delta_{Yk}+\lambda}\right)\nonumber\\&-&\text{Li}_2\left(\frac{\lambda-1}{\delta_{Yk}+\lambda}
\right),
\\
G(x)&=&\log(x ) \log \left(\frac{\delta_{Yk}^2+x_k-
x}{\delta_{Yk}^2+x_k}\right)+\text{Li}_2\left(\frac{x }{\delta_{Yk}^2+x_k}\right),
\\
H(\tau)&=&\text{Li}_2\left(\frac{\tau}{\tau+\delta_{Yk}-1}\right)-\text{Li}_2\left(\frac{\tau}{
\tau-\delta_{Yk}}\right).
\end{eqnarray}
Although each Feynman diagram  is ultraviolet divergent by itself, all the
divergences cancel each other out. This becomes evident when we write
$g_{ZYY}=-2gV=-(g_L+g_R)$. 

We would like to point out that to cross-check our calculation, we made an
alternative evaluation via the Passarino-Veltman method, using the unitary gauge and without any
approximation. The resulting
amplitudes are rather lengthy to be included here, but numerical evaluation showed a nice
agreement between our approximate result and the exact calculation.

The \zd~ decay width is given in the massless outgoing lepton limit
by
\begin{equation}
\Gamma(Z\to \ell^\pm_i \ell^\mp_j)=\frac{m_Z}{24\pi} \left(|F^{ij}_L|^2+ |F^{ji}_L|^2\right).
\label{Zdecaywidth}
\end{equation}
To obtain an estimate for the \zd~branching ratios, we need to analyze the bounds on the mixing
matrix $V^Y$ and the bilepton mass. Below we will
examine the bounds obtained from  the $\ell_j^- \to \ell_i^+
\ell^-_k \ell^-_k$ and $\ell_j\to \ell_i \gamma$ decays together with the muon
anomalous magnetic moment.

\section{Numerical analysis}
\label{analysis}
\subsection{Bounds on the bilepton LFV couplings}

\begin{figure}
 \centering
\includegraphics[width=2.0in]{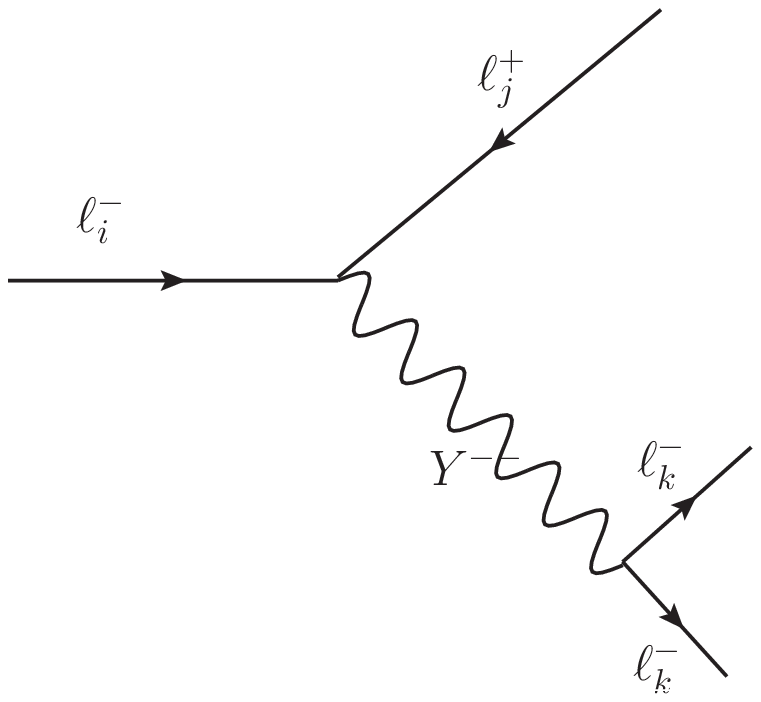}
\caption{Feynman diagrams for the \ldt~ decay at the one-loop level in the \tm.   \label{lito3lk}}
\end{figure}

The three-body decay \ldt~
proceeds at the tree-level through the Feynman diagram of Fig.
\ref{lito3lk}. Its decay width can be obtained straightforwardly in
the limit of massless outgoing leptons. It is given by:

\begin{equation}
\Gamma(\ell_i^-\to  \ell_j^+ \ell_k^- \ell_k^-)=\frac{g^4m_j^5}{3
\,2^{11} \pi^3
m_Y^4}|V^Y_{kk}|^2\left(|V^Y_{ij}|^2+|V^Y_{ji}|^2\right).
\end{equation}

The experimental 90\% C.L. limits  on the LFV decays
$\mu^-\to e^+ e^- e^-$,\cite{Bellgardt:1987du} $\tau^-\to e+ e- e-$,\cite{Aubert:2007pw} and
$\tau^-\to \mu^+
e^- e^-$\cite{Miyazaki:2007zw} translate into the following bounds with 90\% C.L.

\begin{eqnarray}
\frac{|V^Y_{11}|^2\left(|V^Y_{12}|^2+|V^Y_{21}|^2\right)}{m_Y^4}&\le&2.57 \times
10^{-20} \quad {\rm GeV^{-4}},\\
\frac{|V^Y_{11}|^2\left(|V^Y_{13}|^2+|V^Y_{31}|^2\right)}{m_Y^4}&\le&5.14 \times
10^{-12}\quad {\rm GeV^{-4}},\\
\frac{|V^Y_{11}|^2\left(|V^Y_{23}|^2+|V^Y_{32}|^2\right)}{m_Y^4}&\le&5.29 \times
10^{-12}\quad {\rm GeV^{-4}}.
\end{eqnarray}
We used the $\mu$ and $\tau$ mean lifetimes given in Ref. \refcite{pdg}. Following Ref.
\refcite{Liu:1993gy}, we will parametrize the mixing LFV matrix as $V^Y_{ij}\simeq
\delta_ij+\alpha_{ij} e^{i \theta_{ij}}$, where $\alpha_{ij}=-\alpha_{ji}$ and
$\theta_{ij}=-\theta_{ji}$ are  mixing angles and  CP-violating phases, respectively. Therefore we
obtain the followint limits with 90\% C.L.:

\begin{eqnarray}
\label{V12bound1}
\frac{|V^Y_{12}|^2}{m_Y^4}&<&1.28 \times 10^{-20}\quad {\rm GeV^{-4}},\\
\label{V13bound1}
\frac{|V^Y_{13}|^2}{m_Y^4}&<&2.57 \times 10^{-12}\quad {\rm GeV^{-4}},\\
\label{V23bound1}
\frac{|V^Y_{23}|^2}{m_Y^4}&<&2.64 \times
10^{-12}\quad {\rm GeV^{-4}}.
\label{boundslito3lj}
\end{eqnarray}

\begin{figure}
 \centering
\includegraphics[width=3.5in]{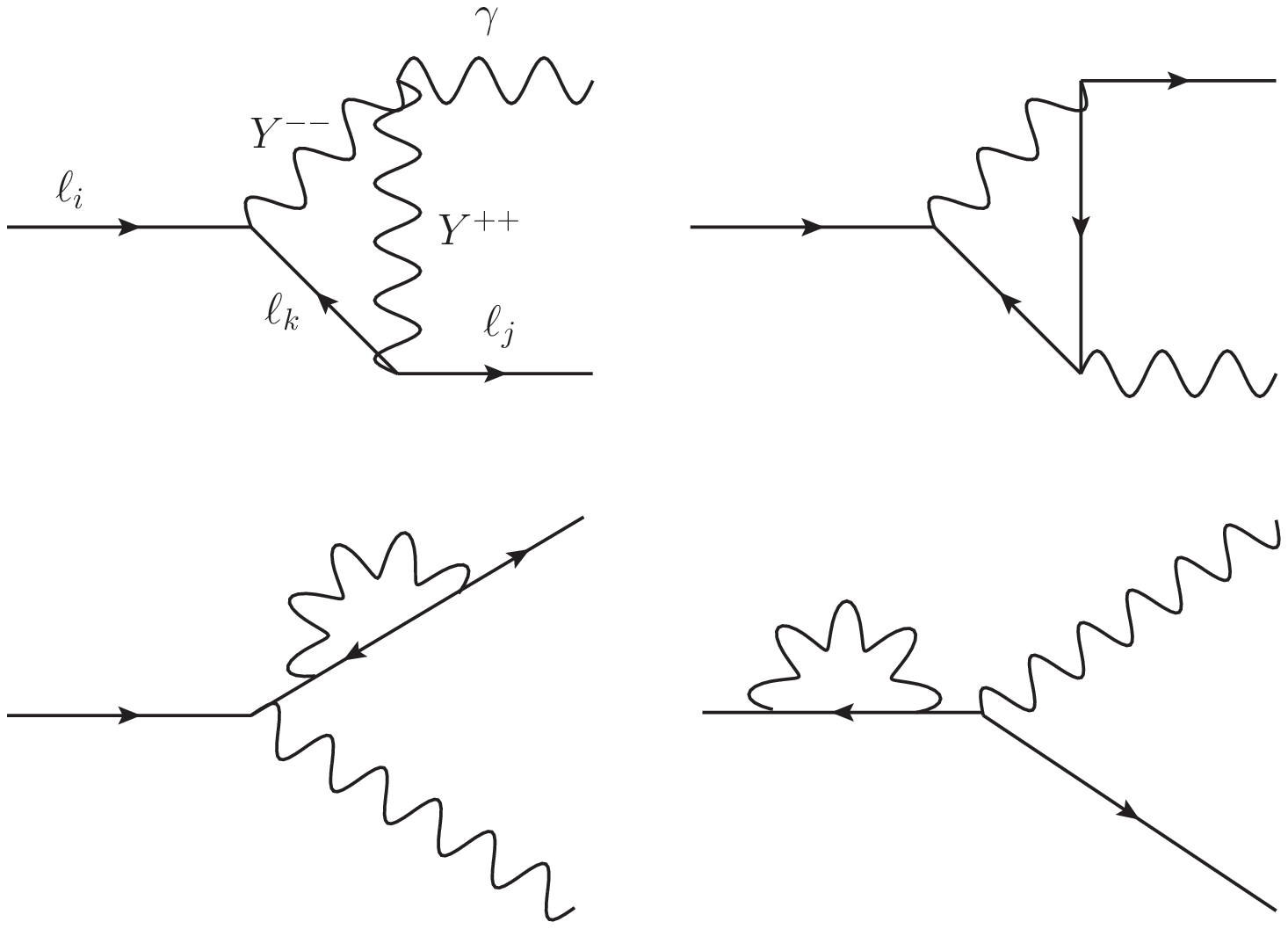}
\caption{Feynman diagrams for the \ld~ decay at the one-loop level
in the \tm.  \label{fig_ljligamma}}
\end{figure}

As far as the $\ell_i\to \ell_j \gamma$ decay, it proceeds via the triangle
Feynman diagrams shown in Fig. \ref{fig_ljligamma}. The bubble
diagrams does not contribute in the limit of massless outgoing
lepton. Using the same scheme described above to obtain the \zd~ amplitude, we obtain the \ld~
amplitude, which can be written as
\begin{equation}
\label{mljtolig} {\cal M}(\ell_i\to \ell_j \gamma)=
\frac{ieq_{\mu}}{m_i+m_j}\,\bar{u}(p_j)
\sigma^{\alpha\mu}\left(f_{A}^{ij} + f_{V}^{ij} \gamma^5\right)
u(p_i) \epsilon_\alpha(p),
\end{equation}
where $q=p_i-p_j$ and $p_{i}$ ($p_j$) are the photon and the
incoming (outgoing) lepton 4-momenta, whereas

\begin{equation}
f_{V,A}^{ij}=\frac{g^2m_i}{2^6\pi^2 m_Y^2}\int^1_0
\int_0^{1-x}\left(\frac{x}{M_1}+\frac{2(1-x)}{M_2}\right)\sum_{k=1}^3\left(V^Y_{ik}V^{Y*}_{kj}\pm
V^{Y*}_{jk}V^Y_{ki}\right)m_k,
\end{equation}
with $M_1=x(1-y\, \xi_i)+(1-x)\xi_k$, $M_2=x(\xi_k-y\,
\xi_i)+(1-x)$ and $\xi_a=m_a^2/m_Y^2$. We have dropped any terms independent of
the internal lepton mass, which cancel due to the unitarity of the
$V^Y$ matrix. Furthermore, if one neglects the lepton masses in the
$M_1$ and $M_2$ coefficients, it follows that
\begin{equation}
\label{fVA}
f_{V,A}^{ij}=\frac{3g^2m_i}{2^6\pi^2
m_Y^2}\sum_{k=1}^3\left(V^Y_{ik}V^{Y*}_{kj}\pm
V^{Y*}_{jk}V^Y_{ki}\right)m_k,
\end{equation}
which coincides with the result previously found in Ref.
\refcite{Liu:1993gy}. In the massless outgoing lepton limit, the \ld~decay width can be written as
\begin{equation}
\Gamma(\ell_i\to l_j
\gamma)=\frac{e^2m_i}{8\pi}\left(|f_V^{ij}|^2+|f_A^{ij}|^2\right).
\end{equation}
Since the sum in Eq. (\ref{fVA})  depends on the internal lepton mass and the $\mu^-\to e^+ e^- e^-$
decay gives a strong constraint on $|V^Y_{12}|$, we can neglect the electron and muon terms. It
follows that the experimental 90\% C.L.  limits on the decays $\mu\to e\gamma$,\cite{Adam:2011ch}
$\tau\to e\gamma$,\cite{Ritt:2006cg} and $\tau\to \mu \gamma$\cite{:2009tk} translate into the
following bounds with 90\% C.L.

\begin{eqnarray}
\label{V13V23b}
\frac{|V^{Y}_{13}|^2|V^Y_{23}|^2}{m_Y^4}&\le&1.29 \times 10^{-21}\quad {\rm GeV^{-4}},\\
\label{V13b}
\frac{|V^{Y}_{13}|^2}{m_Y^4}&\le&2.35 \times 10^{-13}\quad {\rm GeV^{-4}}, \\
\label{V23b}
\frac{|V^{Y}_{23}|^2}{m_Y^4}&\le&3.14 \times 10^{-13}\quad {\rm GeV^{-4}}.
\end{eqnarray}

As far as the lepton anomalous magnetic moment $a_\ell=(g-2)/2$ is concerned,  it can receive new
contributions in the \tm~
from the scalar and gauge sectors. The doubly charged bilepton contributes through triangle
diagrams analogue to those of Fig. \ref{fig_ljligamma}. Its contribution is
given by

\begin{equation}
a_\ell=\frac{g^2m_\ell}{3\,2^3\pi^2 m_Y^2}\sum_{k=1}^3|V^Y_{\ell k}|^2\left(9m_k \cos(2\theta_{\ell
k})+ 7m_\ell
\right).\label{amu331}
\end{equation}
If the flavor mixing matrix $V^Y$ is diagonal, $a_\ell$ becomes
\begin{equation}
a_\ell=\frac{2g^2m_\ell^2}{3\pi^2 m_Y^2},
\end{equation}
which agrees with the result obtained before in Ref. \refcite{Ky:2000ku}.
We also observe that the contribution to $a_\ell$ from the scalar bosons as well as the singly
charged bilepton and the extra neutral gauge boson are subdominant\cite{Ky:2000ku}, so the \tm~
contribution, $a_\ell^{331}$, can be assumed to arise  mainly from the doubly charged bilepton. 
The current experimental limit on $a_\mu$ is not useful to bound the LFV couplings but it can
constrain the bilepton mass $m_Y$. For the theoretical and  experimental values of $a_\mu$, we will
use the most recent data quoted in Ref. \refcite{pdg}. The discrepancy between the theoretical SM
contribution, $a_\mu^{\rm SM}$, and the world average, $a_\mu^{\rm Exp.}$, of experimental
measurements\cite{Bennett:2006fi} is given by:

\begin{equation}
\Delta a_\mu=a_\mu^{\rm Exp.}-a_\mu^{\rm SM}= 255\,(63)\,(49) \times 10^{-11},
\end{equation}
where the $e^-e^+\to \pi^-\pi^+$ data from BABAR were used to evaluate the hadronic contribution to
$a_\mu^{\rm SM}$.\cite{Davier:2009zi} Although this discrepancy is about 3.2 standard deviations,
it is not yet conclusive since there is still a considerable discrepancy between the various
evaluations of the hadronic contribution. If $\Delta a_\mu$ is ascribed to the doubly charged
bilepton, we get the following bound on $m_Y$ with 95 \% C.L.:
\begin{equation}
m_Y\ge 421 \,\, {\rm GeV}.
\end{equation}
Let us assess how this bound compares with other indirect bounds. The very
stringent bound $m_Y>800$ GeV was obtained from muonium-antimuonium conversion.
\cite{Willmann:1998gd} This bound, which  would rule out the minimal
\tm, is based on the assumptions that the $V^Y$ matrix is flavor diagonal and the scalar sector of
the model does not contribute significantly to muonium-antimuonium conversion. Another stringent
bound, $m_Y>750$ GeV, arises from fermion pair production and lepton-flavor violating processes.
\cite{Tully:1999yg} These bounds can be evaded if one considers an extended Higgs sector or less
restrictive assumptions.\cite{Pleitez:1999ix} We will rather
consider a bilepton mass of a few hundreds of GeV to obtain
an estimate of the \zd~ branching ratios.

\subsection{The \zd~branching ratios}
Due to the unitarity of $V^Y$, i.e. $\sum_k V^Y_{ik}V^{*Y}_{kj}=\delta_{ij}$, and neglecting
imaginary phases, we can
write

\begin{eqnarray}
{\rm BR}(Z\to \ell_i^\mp
\ell_j^\pm)&=& \lambda
\Big|V^{Y}_{i2} V^Y_{j2}\left(I(x_2,x_Y)-
I(x_1 , x_Y)\right)\nonumber\\&+&V^Y_{i3} V^Y_{j3}\left(I(x_3 , x_Y)-I(x_1 , x_Y)\right)\Big|^2,
\end{eqnarray}
with $
 \lambda=\frac{m_Z}{12\pi\Gamma_Z}\left(\frac{g^3}{2^6 \pi^2c_W}\right)^2$.
Notice that according to Eq. (\ref{V12bound1}), $|V^Y_{12}|$ is strongly constrained, so one can
write
\begin{eqnarray}
{\rm BR}(Z\to e^\mp
\mu^\pm)&\simeq& \lambda \left|V^Y_{13}\right|^2\left| V^Y_{23}\right|^2\left|I(x_3 ,
x_Y)-I_1(x_3,x_Y)\right|^2,\\
{\rm BR}(Z\to e^\mp
\tau^\pm)&\simeq&\lambda\left|V^{Y}_{13}\right|^2\left|
I(x_3,x_Y)-I(x_1 , x_Y)\right|^2,\\
{\rm BR}(Z\to \mu^\mp
\tau^\pm)&\simeq&\lambda\left|
V^Y_{23}\right|^2\left|I(x_3 , x_Y)-I(x_2 , x_Y)\right|^2.
\end{eqnarray}

Numerical evaluation together with the bounds (\ref{V13V23b})-(\ref{V23b}) give the
upper bounds on the LFV $Z$ decays shown in Table \ref{BRbounds} for $m_Y=100$ GeV and $m_Y=500$
GeV. While the loop amplitude magnitude decreases for larger values of $m_Y$, the bounds on the LFV
matrix elements $V^Y_{ij}$ loosen up. This explains the fact that the bounds on the LFV $Z$
decays are slightly weaker for larger $m_Y$. However, the bounds
obtained for $m_Y=1000$ GeV are of similar order of magnitude than those obtained for $m_Y=500$ GeV.
Our results indicate that the bilepton mediated $Z$ decays would be far from the reach of detection,
though the corresponding branching ratios are of the same order of magnitude than in 
other SM extensions. For instance, in the framework of the Zee-model it was found that ${\rm
BR}(Z\to e^\mp\mu^\pm)<4.2 \times 10^{-16}$, ${\rm
BR}(Z\to e^\mp\tau^\pm)<1.1 \times 10^{-8}$, and ${\rm
BR}(Z\to \mu^\mp\tau^\pm)<1.6 \times 10^{-12}$ 
 for typical values of the model parameters;\cite{Ghosal:2001ep} also, in the SM enlarged with
massive neutrinos with masses of the order of a few dozens of GeV, the upper bound
$BR(Z\to\mu^\mp\tau^\pm)\lesssim 10^{-11}$ was obtained.\cite{Illana:2000ic} 

\begin{table}
\tbl{Upper limit with 90\%C.L. on the bilepton contribution to LFV $Z$ decays.
The bounds given in  Eqs. (\ref{V13V23b})-(\ref{V23b}) were employed.}
{\begin{tabular}{|c|c|c|}
\hline
BR&$m_Y=100$ GeV&$m_Y=500$ GeV
\\ \hline
${\rm BR}(Z\to e^\mp
\mu^\pm)$&$3.2 \times 10^{-20}$&$1.1\times^{-19}$\\
\hline
${\rm BR}(Z\to e^\mp
\tau^\pm)$&$6.37 \times 10^{-11}$&$4.57 \times 10^{-10}$ \\ \hline
${\rm BR}(Z\to \mu^\mp \tau^\pm)$&$6.56\times 10^{-11}$&$2.26\times 10^{-10}$ \\ \hline
\end{tabular}\label{BRbounds}}
\end{table}

\section{Final remarks}
\label{conclusion}
We have investigated the \zd~decay in the framework of the minimal 331 model. We focused on the
contributions mediated by the doubly charged bilepton
and estimated the bounds on the LFV bilepton couplings from the experimental
constraints on the decays \ld~ and \ldt. Our results indicate that these
contributions seem to be far from experimental detection. The smallness of the decay rates can be
explained mainly from the fact that the LFV bilepton couplings  are strongly constrained by
current experimental data.

\section*{Acknowledgements}
We acknowledge support from Conacyt and SNI (M\'exico). Support from VIEP-BUAP
is also acknowledged.

\end{document}